\documentclass[preprint,onecolumn,11pt,nofootinbib]{revtex4}
\usepackage{braket}
\usepackage{float}
\usepackage{graphics,epsfig}
\usepackage{bm}
\usepackage{slashed}
\usepackage{graphicx}
\usepackage{amsmath}
\usepackage{amsfonts}
\usepackage{amssymb}
\usepackage{color}%
\usepackage{dcolumn}
\usepackage[
            pdfstartview=FitH,
            bookmarksnumbered=true,
            bookmarksopen=true,
            colorlinks,
            linkcolor=blue,
            anchorcolor=green,
            citecolor=red
            ]{hyperref}
\usepackage{enumerate}
\providecommand{\U}[1]{\protect\rule{.1in}{.1in}}

\def\A0{A^{(0)}}

\begin{document}
\baselineskip=0.6 cm \title{Holographic complexity for nonlinearly charged Lifshitz black holes}
\author{Kai-Xin Zhu$^{1,2}$}
\thanks{E-mail address: kaixinzhuabc@gmail.com}
\author{Fu-Wen Shu$^{1,2,3}$}
\thanks{E-mail address: shufuwen@ncu.edu.cn}
\author{Dong-Hui Du$^{1,2}$}
\thanks{E-mail address: donghuiduchn@gmail.com}
\affiliation{
$^{1}$Department of Physics, Nanchang University, Nanchang, 330031, China\\
$^{2}$Center for Relativistic Astrophysics and High Energy Physics, Nanchang University, Nanchang 330031, China\\
$^{3}$Center for Gravitation and Cosmology, Yangzhou University, Yangzhou, China }


\begin{abstract}
Using ``complexity=action'' proposal we study the late time growth rate of holographic complexity for nonlinear charged Lifshitz black hole with a single horizon or two horizons. As a toy model, we consider two kinds of such black holes: nonlinear charged Lifshitz black hole and nonlinear logarithmic charged Lifshitz black hole. We find that for the black hole with two horizons, the action growth bound is satisfied. But for the black hole with a single horizon, whether the Lloyd bound is violated depends on the specific value of dimensionless coupling constants $\beta_{1},\beta_{2}$, spacetime dimension $D$ and dynamical exponent $z$.
\end{abstract}

\maketitle
\newpage
\vspace*{0.2cm}

%
\section{INTRODUCTION }
Many significant concepts emerged in the exploration of the unification of general relativity and quantum mechanics. One of them is the work of holographic principle \cite{tHooft:1993dmi,Susskind:1994vu}, which states that degrees of freedom of a higher dimensional gravitational system can be characterized by those of a lower dimensional quantum system. This principle is widely regarded as a fundamental principle of quantum gravity with the discovery of AdS (Anti-de Sitter)/CFT (Conformal field theory) correspondence \cite{Maldacena:1997re}. Based on this duality, the so-called holographic dictionary is proposed \cite{Gubser:1998bc,Witten:1998qj} and it offers us a new way to calculate the physical quantities on the field theory side.  In addition, increasing evidences  show that it may play a crucial role in the understanding the nature of space-time \cite{VanRaamsdonk:2010pw,Swingle:2009bg}. In particular, as implied by the holographic entanglement entropy \cite{Ryu:2006bv,Hubeny:2007xt}, there is a fundamental connection between quantum information theory and gravitational physics \cite{Czech:2017ryf,Bao:2017qmt,Bao:2017qcc,Du:2019vwh,Du:2019emy,Gan:2017qkz,Chen:2018vkw,Chen:2018ywy,Chen:2018ody}. The quantum complexity is another significant concept of the connection and thus has attracted a lot of attention in the past few years. It is defined by the minimal number of quantum gates needed to build a target state from a reference state within small tolerance $\epsilon$. Two proposals on holographic duality of quantum computational complexity have emerged.
First, the complexity = volume (CV) conjecture \cite{Susskind:2014rva,Stanford:2014jda}, which states that complexity of the boundary
state was proportional to the maximum volume of codimension one hypersurface bounded by the CFT slices,
\begin{equation}
\mathcal{C_{V}}\sim\frac{\mathcal{V(B)}}{G_{N}\ell},
\end{equation}
where $G_{N}$ is the Newton's constant and $\ell$ is a arbitrary length scale. In order to avoid this unclear length scale appearing in the CV conjecture, the conjecture of complexity = action (CA) was proposed \cite{Brown:2015bva,Brown:2015lvg}, which states that complexity on
 the boundary CFT is related to the on-shell action of the Wheeler-DeWitt(WDW) patch in the bulk, which is defined as the domain of dependence of  Cauchy surface that is anchored on the boundary state
\begin{equation}
\mathcal{C_{A}} =\frac{ I_ {WDW}}{\pi \hbar}.
\end{equation}
 Both these conjectures  favor a statement that there exists a bound on the growth rate of complexity for neutral black holes which is known as the Lloyd bound \cite{Lloyd}. It suggests that the rate of computation is bounded by the total energy of the system
\begin{eqnarray}
\frac{d\mathcal{C}}{dt}\leq\frac{2M}{\pi\hbar},
\end{eqnarray}
where $M$ is the mass of the black hole. For charged or rotated AdS black hole the bound is
 modified \cite{Cai:2016xho} as,
\begin{eqnarray}
\frac{d\mathcal{C}}{dt}=\frac{1}{\pi\hbar}\biggl[(M-\Omega J-\mu Q)_{+}-(M-\Omega J-\mu Q)_{-}\biggr]
\end{eqnarray}
where $\mu$ and $\Omega$ are black hole chemical potential and angular velocity, $J$ and $Q$ are black hole angular momentum and electric charge, ``+'' and ``-'' stand for outer horizon and inner horizon of the black hole, respectively. The validity of this upper bound in different types of black hole have been checked in many literatures\cite{Pan:2016ecg,Cano:2018aqi,Ovgun:2018jbm,Meng:2018vtl,Feng:2018sqm,Jiang:2018tlu,Jiang:2018pfk}. However, there also have many cases that violate the Lloyd bound \cite{Ghaffarnejad:2018prc,Nagasaki:2018csh,Fan:2019aoj}. The violation of the Lloyd bound in some systems implies that  the CV or CA conjecture should be modified in these systems.  Hence, to test how common the CV or CA conjecture is deserves further study.

For this sake, in this paper we focus on the holographic complexity in the non-realistic system, as all the aforementioned cases are restricted to the relativistic systems. Specifically, we discuss whether spacetime anisotropy will have an influence on the validity of the action growth bound of holographic complexity, under the framework of Lifshitz holography, which was first addressed in \cite{Kachru:2008yh,Taylor:2008tg}. In the Lifshitz system, the time and space coordinates rescale with different weight
\begin{eqnarray}
t\mapsto\lambda^{z}t,\quad\quad \vec{x}\mapsto\lambda\vec{x}
\end{eqnarray}
where the constant $z$ is called the dynamical critical exponent. In the anisotropic systems, the Lloyd bound was shown to be violated at late
times for some cases, such as in the non-commutative geometry \cite{Couch:2017yil}, and the Einstein-Maxwell-Dilaton theory defined in the Lifshitz and hyperscaling violating geometries \cite{Swingle:2017zcd,An:2018xhv,Alishahiha:2018tep,Alishahiha:2019lng}. However, all of these cases are discussing about the neutral black hole. The holographic complexity in the charged Lifshitz black hole, which has much richer structure, is still lack.

In this paper we consider a toy model. We use the CA conjecture to study the holographic complexity of the nonlinearly charged Lifshitz black hole in the Einstein-Proca-Maxwell model. Such non-trivial black hole solution in the Einstein-Proca-Maxwell gravity was obtained in \cite{Pang:2009pd}, and later was developed in \cite{Alvarez:2014pra,Dehghani:2011tx}. Related thermodynamics and CV duality have been investigated in
\cite{Dehghani:2013mba,Ayon-Beato:2015jga,Liu:2014dva}. More specifically, we consider the Einstein-Proca-Maxwell model in Ref \cite{Alvarez:2014pra}. One feature of this Lifshitz system is that the dynamical critical exponent $z$ could be choose arbitrary for $z>1$ ($z\leq1$ is not allowed in our model \cite{Alvarez:2014pra} because Proca field is only possible for $z>1$). With different value of dimensionless coupling constants $\beta_{1},\beta_{2}$, we could get the charged black hole with a single horizon or two horizons. In this way, we could check the validity of the general charged black hole with two horizons and the more specific case where the black hole only has a single horizon, as what did in the relativistic system in \cite{Cai:2017sjv,Meng:2018vtl}. Interestingly, if we choose the dynamical exponent $z=D-2$, charged black hole will become logarithmic decay supported by a specific logarithmic electrodynamics. So we can check whether the holographic complexity is continuous for $z$, in other words, whether the logarithmic behavior will influence the holographic complexity.

Our results show that the nonlinear charged Lifshitz black hole with two horizons the upper bound is always satisfied regardless of the value of parameters of the model, while for those with a single horizon, the Lloyd bound could be violated depending on different value of the model. We also find in the side of gravity when $z=D-2$ the Lifshitz black hole will yields logarithmic decay behavior, but on the side of holographic complexity it does not yield any specific influence.

This paper is organized as follows: In section II, we make a basic setup of the model, from which we obtain the black hole solutions for two different cases: $z\neq D-2$ and $z=D-2$.  Thermodynamics of  these two black holes is investigated respectively. We then calculate action growth of nonlinearly charged Lifshitz black holes for $z\neq D-2$ in section III. In section IV we calculate action growth of nonlinearly logarithmic charged Lifshitz black holes where $z=D-2$. The conclusions and discussions will be presented in section V.

\section{Nonlinearly charged Lifshitz black holes}

In this section we consider the charged Lifshitz black holes for any exponent $z>1$ with action \cite{Alvarez:2014pra}:
\begin{eqnarray}\label{action111}
I=\int{d}^Dx\sqrt{-g}\biggl[\frac1{2\kappa}(R-2\lambda)
-\frac{1}{4}H_{\mu\nu}H^{\mu\nu}-\frac{1}{2}m^2B_{\mu}B^{\mu}-\frac12P^{\mu\nu}F_{\mu\nu}+\mathcal{H}(P)
\biggr],
\end{eqnarray}
where $B_{\mu}$ is a massive vector field (Proca field) with strength $H_{\mu\nu}=\partial_{u}B_{\nu}-\partial_{\nu}B_{\mu}$, $m$ is its mass. Usually the last
two terms in the action are the functions of $F=\frac{1}{4}F_{\mu\nu}F^{\mu\nu}$ and describe the nonlinear behavior of the electromagnetic field $A_{\mu}$ with field
strength $F_{\mu\nu}=\partial_{u}A_{\nu}-\partial_{\nu}A_{\mu}$. Using Legendre transformation $\mathcal{H}(P)$ this Lagrangian could be rewriten as a
function of conjugate antisymmetric tensor $P^{\mu\nu}$ and electromagnetic field $A_{\nu}$ \cite{Salazar:1987ap}, where $\mathcal{H}(P)$ is the so-called
structural function depending on the invariant $P=\frac{1}{4}P_{\mu\nu}P^{\mu\nu}$. After variation of the action with respect to $P^{\mu\nu}$, the strength
field of the original nonlinear electromagnetic field can be express as
\begin{equation}\label{Legendre}
F_{\mu\nu}=\mathcal{H}_{P}P_{\mu\nu},
\end{equation}
where $\mathcal{H}_{P}=\partial\mathcal{H}/{\partial P}$. When $\mathcal{H}(P)=P$ it reduces to standard linear Maxwell theory. In general, the well-behaved
nonlinear electrodynamics need express as a polynomial of the invariants $F_{\mu\nu}$ \cite{Brynjolfsson:2009ct,Balasubramanian:2010uw}.
The Lifshitz black holes with generic dynamical exponent $z$ can be obtained by choosing the following structural function \cite{Alvarez:2014pra}:
\begin{eqnarray}\label{structural function}
\mathcal{H}(P)&=&-
\frac{[2z^2-Dz+2(D-2)]}{2\kappa l^2}\beta_1\sqrt{-2l^2P}
-\frac{(z-1)(z-D+2)^2}{\kappa z}{\beta_1}^2P \nonumber\\\nonumber\\
&&{}+\frac{(D-2)^2}{2\kappa l^2}\beta_2\left(-2l^2P\right)^{\frac{z}{2(D-2)}},
\end{eqnarray}
where $\beta_1$ and $\beta_2$ are dimensionless coupling constants. Varying the action \eqref{action111} with respect to $B_{\mu}$, $A_{\mu}$ and $g_{\mu\nu}$, one gets the equations of motion, respectively, for Proca field, electromagnetic field and metric filed
\begin{eqnarray}\label{equations of motion(1)}
\nabla_\mu H^{\mu\nu}&=&m^2B^\nu;\\
\nabla_\mu P^{\mu\nu}&=&0;\label{vary A}
\end{eqnarray}
\begin{eqnarray}\label{equations of motion(2)}
G_{\mu\nu} + \lambda g_{\mu\nu} &=& \kappa \biggl[
H_{\mu\alpha}H_\nu^{~\alpha}-\frac14g_{\mu\nu}H_{\alpha\beta}H^{\alpha\beta} +
m^2\left(B_\mu B_\nu-\frac12g_{\mu\nu}B^\alpha B_\alpha\right) \nonumber\\
&&{}\quad+\mathcal{H}_P P_{\mu\alpha}P_\nu^{~\alpha}-g_{\mu\nu}(2P\mathcal{H}_P-\mathcal{H}) \biggr].
\end{eqnarray}
These equations admit the following charged Lifshitz black hole solution \cite{Alvarez:2014pra}

\begin{eqnarray}
\label{metric}
ds^2&=&-\frac{r^{2z}}{l^{2z}}f(r)dt^2+\frac{l^2}{r^2}\frac{dr^2}{f(r)}+r^2d\vec{x}^2,\\
P_{\mu\nu}&=&2\delta_{[\mu}^t\delta_{\nu]}^r\frac{Q}{l^{D-2}}\left(\frac{l}{r}\right)^{D-z-1},\label{antisymmetric tensor}\\
B_t(r)&=&\sqrt{\frac{z-1}{z \kappa}}\left(\frac{r}{l}\right)^zf(r),\label{proca field}
\end{eqnarray}
The expression for $f(r)$ depends on the parameters $z$ and $D$ and its explicit form will be given later.

\subsection{Nonlinearly charged Lifshitz black holes with exponent $z>1$ and $z\neq D-2$}

For nonlinearly charged Lifshitz black hole with exponent $z>1$ and $z\neq D-2$, we have\cite{Alvarez:2014pra}
\begin{eqnarray}
f(r)&=&1-M_1\left(\frac{l}{r}\right)^{D-2}+M_2\left(\frac{l}{r}\right)^z,\label{gravitational potential}\\
\lambda&=&-\frac{z^2+(D-3)z+(D-2)^2}{2l^2},\\
M_1&=&\beta_1\frac{|Q|}{l^{D-3}}, \quad
M_2=\beta_2\left(\frac{|Q|}{l^{D-3}}\right)^{\frac{z}{D-2}},\label{integration constant}
\end{eqnarray}
where the integration constants $M_{1}$ and $M_{2}$ are related to the mass of the black hole, and $Q$ is related to the electric charge.

Before calculating the holographic complexity, we investigate the thermodynamics of the nonlinearly charged Lifshitz black holes with exponent $z>1$ and $z\neq D-2$.
The entropy per unit volume of the horizon is given, as usual, by the Bekenstein-Hawking formula
\begin{equation}\label{entropy}
S=\frac{2\pi r_{+}^{D-2}}{\kappa}.
\end{equation}
The temperature of the event horizon can be obtained by using the standard Wick-rotation method, yielding the result
\begin{equation}\label{temperture}
T=\frac{r_{+}^{z+1}}{4\pi l^{z+1}}f_{r=r_{+}}^{'}=\frac{1}{4\pi l}\left(\frac{r_{+}}{l}\right)^{z}\biggl[(D-2)M_{1}\left(\frac{l}{r_{+}}\right)^{D-2}-zM_{2}\left(\frac{l}{r_{+}}\right)^{z}\biggr].
\end{equation}
The electric charge density reads
\begin{equation}\label{charge}
Q_{e}=\frac{1}{\Omega_{k}}\int d\Omega_{k}r^{D-2}n^{\mu}P_{\mu\nu}u^{\nu},
\end{equation}
where $n^{\mu}$ and $u^{\nu}$ are the unit spacelike and timelike normals to a sphere of radius $r$
\begin{eqnarray}
u^{\nu}=\frac{1}{\sqrt{-g_{tt}}}dt=\frac{l^{z}}{r^{z}\sqrt{f}}dt,
n^{\mu}=\frac{1}{\sqrt{g_{rr}}}dr=\frac{r\sqrt{f}}{l}dr.
\end{eqnarray}
Substituting the above formulae, one obtains
\begin{equation}\label{charge}
Q_{e}=Q.
\end{equation}
As a consequence, similar as \cite{Liu:2014dva}, the electric potential reads
\begin{eqnarray}
A_{t}(r)&=&\frac{2z^{2}-Dz+2(D-2)}{2z\kappa}M_{1}\frac{l^{D-3}}{Q}(\frac{r}{l})^{z}-\frac{(z-1)(z-D+2)}{z\kappa}M_{1}^{2}\frac{l^{D-3}}{Q}(\frac{r}{l})^{z-D+2}\nonumber\\
&&-\frac{z}{2\kappa}M_{2}\frac{l^{D-3}}{Q}(\frac{r}{l})^{D-2},
\end{eqnarray}
\begin{equation}\label{charge potential}
\begin{aligned}
\Phi_{e}=-A(r_{+})&=\frac{l^{D-3}}{2\kappa Q }\biggl[-(D-2)M_{1}\left(\frac{r_{+}}{l}\right)^{z}+zM_{2}\left(\frac{r_{+}}{l}\right)^{D-2}+\frac{2(z-1)(z-D+2)}{z}M_{1}M_{2}\biggr].
\end{aligned}
\end{equation}
On the other hand, the mass of the black hole can be computed through the quasilocal method as described in Refs\cite{Dehghani:2013mba,Gim:2014nba}
with the following relation
\begin{equation}\label{mass}
\tilde{Q}(\xi)=\int_{B} d^{D-2}x_{\mu\nu}\left(\Delta K^{\mu\nu}(\xi)-2\xi^{[\mu}\int^{1}_{0}ds\Theta^{\nu]}\right),
\end{equation}
where $\Theta^{\nu}$ is the surface term, $\Delta K^{\mu\nu}(\xi)\equiv K_{s=1}^{\mu\nu}(\xi)-K_{s=0}^{\mu\nu}(\xi)$ denotes the variation of the Noether potential
from the vacuum solution to the black hole, and $dx_{\mu\nu}$ represents the integration over the codimension-two boundary $B$. As shown in (\ref{gravitational potential}),
the solution depends continuously on the integration constants $M_{1}$ and $M_{2}$, and the Lifshitz black hole has the mass as the only charge. Hence we can introduce the
parameter $sM_{1}$ with $s\in[0,1]$  in the space of solutions to define the conserved charge in the interior region (not in the asymptotic region)  \cite{Barnich:2003xg,Barnich:2004uw,Wald:1999wa,Barnich:2007bf}. In our case (\ref{action111}), the involved quantities are \cite{Bravo-Gaete:2015iwa}
\begin{eqnarray}\label{surface term }
\Theta^{\mu}&=&2\sqrt{-g}\biggl[P^{\mu(\alpha\beta)\gamma}\nabla_{\gamma}\delta g_{\alpha\beta}-\delta g_{\alpha\beta}\nabla_{\gamma}P^{\mu(\alpha\beta)\gamma}
+\frac{1}{2}\frac{\partial\mathcal{L}}{\partial (\partial_{\mu}A_{\nu})}\delta A_{\nu}+\frac{1}{2}\frac{\partial\mathcal{L}}{\partial (\partial_{\mu}B_{\nu})}\delta B_{\nu}\biggr],\\
\label{noether potential}
K^{\mu\nu}&=&\sqrt{-g}\biggl[2P^{\mu\nu\rho\sigma}\nabla_{\sigma}\xi_{\sigma}-4\xi_{\sigma}\nabla_{\rho}P^{\mu\nu\rho\sigma}-\frac{1}{2}\frac{\partial\mathcal{L}}{\partial
(\partial_{\mu}A_{\nu})}\xi^{\sigma}A_{\sigma}-\frac{1}{2}\frac{\partial\mathcal{L}}{\partial (\partial_{\mu}B_{\nu})}\xi^{\sigma}B_{\sigma}\biggr],
\end{eqnarray}
where $P^{\mu\nu\rho\sigma}=\frac{\partial\mathcal{L}}{\partial R_{\mu\nu\rho\sigma}}$. In the present case, the last expression becomes
\begin{eqnarray}\label{surface term for my case}
\int^{1}_{0}ds\Theta^{r}&=&\frac{l^{D-3}}{2\kappa}\biggl[2\left(\frac{r}{l}\right)^{z}M_{1}-2\left(\frac{r}{l}\right)^{D-2}M_{2}-\frac{2(z-1)(z-D+2)}{z+D-2}M_{1}M_{2}\biggr],\\
\Delta K^{rt}&=&\frac{l^{D-3}}{2\kappa}\biggl[-2\left(\frac{r}{l}\right)^{z}M_{1}+2\left(\frac{r}{l}\right)^{D-2}M_{2}+\frac{2(z-1)(z-D+2)}{z}M_{1}M_{2}\biggr].\label{noether potential for my case}
\end{eqnarray}
From these expressions, we obtain the mass of the nonlinearly charged Lifshitz black holes with exponent $z>1$ and $z\neq D-2$
\begin{equation}\label{mass for my case}
M=\frac{l^{D-3}}{\kappa}\Omega_{D-2}\frac{(z-1)(z-D+2)}{z}\frac{D-2}{z+D-2}M_{1}M_{2},
\end{equation}
and it is easy to check that the first law holds
\begin{equation}\label{first law}
dM=TdS+\Phi_{e} dQ_{e}.
\end{equation}
The Smarr formula turns out to be
\begin{equation}\label{Smarr formula}
M=\frac{D-2}{z+D-2}(TS+\Phi_{e}Q_{e}).
\end{equation}
It is worth mentioning when the structural coupling constant $\beta_{1}$ vanishes and $\beta_{2}$ is fixed, the last two terms in the action (\ref{action111})
becomes power law. This kind of power-law Lagrangian has been studied in detail in Ref \cite{Dehghani:2013mba} where they get the same Smarr formula.
On the other hand, when $z=2(D-2)$ structural function  reduced to Maxwell action in Ref \cite{Pang:2009pd}, its thermodynamical properties are
given in \cite{Bravo-Gaete:2015iwa}. Because the structural coupling constant $\beta_{1}$ vanishes, the black hole mass in both cases turns to zero.

It is important to note that the values and signs of the structural coupling constants $\beta_{1},\beta_{2}$ are not limited. This makes it possible to encode
different kinds of black holes, each black hole is associated with a specific electrodynamics behavior as in \cite{Alvarez:2014pra}. We will divide it into
single horizon and two horizons and calculate their holographic complexity, respectively, in section III.

\subsection{Nonlinearly charged logarithmic Lifshitz black hole for exponent $z=D-2$ }

When $z=D-2$ there is a double root, gravitational potential need to be modified with a logarithmic behavior and the resulted black hole becomes a logarithmic charged Lifshitz black holes. To be more explicitly, in this case the gravitational potential (\ref{gravitational potential})  becomes a double multiplicity for the decay power $D-2$, and the above solution can be obtained by redefining the integration constants $(M_{1},M_{2})\mapsto((z-D+2)M_{1},M_{2}-M_{1})$, then the
gravitational potential becomes \cite{Alvarez:2014pra}
\begin{equation}\label{log gravitional potential}
f(r)=1-\left(\frac{l}{r}\right)^{D-2}\left[M_1\ln\left(\frac{r}{l}\right)-M_2\right].
\end{equation}
Meanwile the structural constants $(\beta_{1},\beta_{2})$ should change to $((z-D+2)\beta_{1},\beta_{2}-\beta_{1})$, then the structural function (\ref{structural function})
 is rewriten as
\begin{eqnarray}
\mathcal{H}(P)&=&\frac1{2\kappa l^2}\left\{
\left[(D-2)\ln\sqrt{-2l^2P}-3D+8\right]\beta_1+(D-2)^2\beta_2\right\}\sqrt{-2l^2P}
\nonumber\\\nonumber\\
&&{}-\frac{(D-3)}{(D-2)\kappa }{\beta_1}^2P.\label{eq:Hlog}
\end{eqnarray}
The integration constants can be expressed as
\begin{equation}\label{integration constant 1}
M_1=\beta_1\frac{|Q|}{l^{D-3}}, \quad
M_2=\left(\beta_2+\frac{\beta_1}{D-2}\ln\frac{|Q|}{l^{D-3}}\right)\frac{|Q|}{l^{D-3}}.
\end{equation}
Note that this logarithmic behavior is the limit of $z\rightarrow D-2$ by apllying the $\ln(x)\equiv\lim_{\alpha\rightarrow0}\frac{x^\alpha-1}{\alpha}$ to
(\ref{gravitational potential}), other example for logarithmic black hole is produced by using higher curvature gravity \cite{AyonBeato:2009nh,Lu:2011zk,Alishahiha:2011yb}.

Now let us turn to the thermodynamics of the nonlinearly charged Lifshitz black hole with exponent $z=D-2$. From (\ref{temperture}) we obtain the temperature
\begin{equation}\label{temperturelog}
T=\frac{r_{+}^{D-1}}{4\pi l^{D-1}}f_{r=r_{+}}^{'}=\frac{1}{4\pi l}\biggl\{(D-2)\biggl[M_{1}\ln\left(\frac{r_{+}}{l}\right)-M_{2}\biggr]-M_{1}\biggr\}.
\end{equation}
The electric potential is
\begin{equation}
A_{t}(r)=\frac{1}{2\kappa}\frac{l^{D-3}}{Q}\biggl\{(D-2)\biggl[M_{1}\ln(\frac{r}{l})-M_{2}\biggr](\frac{r}{l})^{D-2}+\frac{2(D-3)}{D-2}M_{1}(\frac{r}{l})^{D-2}
-\frac{2(D-2)}{D-3}M_{1}^{2}\ln(\frac{r}{l})-M_{1}\frac{r}{l}^{D-2}\biggr\},
\end{equation}
\begin{equation}\label{electric potential log}
\Phi_{e}=-A_{t}(r_{+})=\frac{l^{D-3}}{2\kappa Q}\biggl\{-(D-2)\biggl[M_{1}\ln\left(\frac{r_{+}}{l}\right)-M_{2}\biggr]\left(\frac{r_{+}}{l}\right)^{D-2}+M_{1}\left(\frac{r_{+}}{l}\right)^{D-2}
+\frac{2(D-3)}{D-2}M_{1}M_{2}\biggr\}.
\end{equation}
The surface term and the Noether potential could be calculated by using (\ref{surface term }) and (\ref{noether potential})
\begin{eqnarray}\label{surface term for my case log}
\int^{1}_{0}ds\Theta^{r}&=&\frac{l^{D-3}}{2\kappa}\biggl\{2\biggl[M_{1}\ln\left(\frac{r}{l}\right)-M_{2}\biggr]\left(\frac{r}{l}\right)^{D-2}-\frac{D-3}{D-2}M_{1}M_{2}\biggr\},\\
\label{noether potential for my case log}
\Delta K^{rt}&=&\frac{l^{D-3}}{2\kappa}\biggl\{-2\biggl[M_{1}\ln\left(\frac{r}{l}\right)-M_{2}\biggr]\left(\frac{r}{l}\right)^{D-2}+\frac{2(D-3)}{D-2}M_{1}M_{2}\biggr\}.
\end{eqnarray}
Now using (\ref{mass}), one finally obtains
\begin{equation}\label{mass for my case log}
M=\frac{l^{D-3}}{2\kappa}\Omega_{D-2}\frac{D-3}{D-2}M_{1}M_{2}.
\end{equation}
Again we can easily check that the first law holds
\begin{equation}
dM=TdS+\Phi_{e} dQ_{e},
\end{equation}
and the Smarr formula is given by
\begin{equation}\label{Smarr formular log}
M=\frac{1}{2}(TS+\Phi_{e}Q_{e}).
\end{equation}
Actually, substituting $z=D-2$ into (\ref{Smarr formula}) we can get the same result, which means that the logarithmic behavior does not effect the Smarr formula.

\section{ Action growth in  nonlinearly charged Lifshitz black hole with exponent $z>1$ and $z\neq D-2$ }
In this section we would like to calculate the action growth rate in the nonlinearly charged Lifshitz black hole with exponent $z>1$ and $z\neq D-2$ and to see if the action growth bound holds or not. Our computation is based on the complexity=action(CA) conjecture \cite{Lehner:2016vdi} and the discussion is divided into two cases: the single-horizon case and the two-horizon case.
\subsection{ The case with a single horizon}
According to the CA conjecture, the complexity is proportional to the on-shell action in the WDW patch of some time slice. The total action in presence of unsmooth boundaries is given by
\begin{eqnarray}
I_{tot}&=&\int_{\mathcal{M}}d^{D}x\sqrt{-g}(\mathcal{L}_{bulk})\nonumber\\
&&+\frac{1}{\kappa}sign(\Sigma)\int_{\Sigma}d^{D-1}x\sqrt{\mid h\mid}K+\frac{1}{\kappa}sign(N)\int_{N}d^{D-2}x\sqrt{\sigma}\eta\nonumber\\
&&+\frac{1}{\kappa}sign(\Sigma')\int_{\Sigma'}d\lambda d^{D-2}\theta\sqrt{\gamma}\chi+\frac{1}{\kappa}sign(B)\int_{B}d^{D-2}x\sqrt{\sigma}a.
\end{eqnarray}
Here $\Sigma$ is spacelike or timelike boundary, $\Sigma'$ is null boundary and $K$ is the Gibbons-Hawking term.
$\eta$ is the Hayward joint term \cite{Hayward:1993my}, while $a$ is the null joint term.
 $\chi$ measures the failure of $\lambda$ to be an affine parameter on the null generators. Here we choose affine parametrization
 so that the null boundary has no contribution. The signatures $sign(\Sigma), sign(N), sign(\Sigma'),sign(B)$ are determined by the additivity rules.
\begin{figure}[h]
\begin{center}
\includegraphics[width=.90\textwidth]{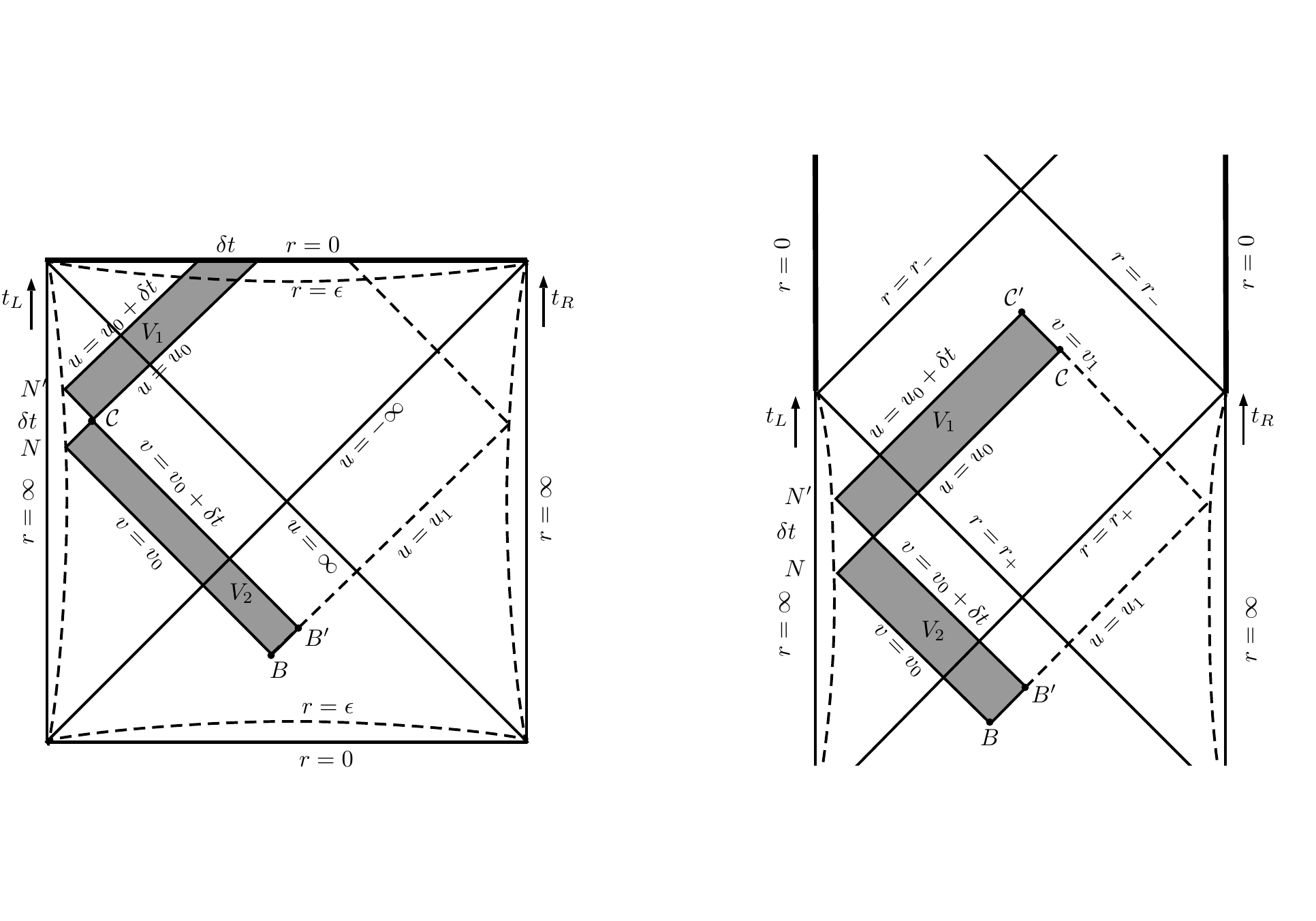}
\end{center}
  \caption{A WDW patch and its change due to an infinitesimal time shift $\delta t$ at the left boundary, for a charged Lifshitz black hole with a single horizon
  (left panel), and for a charged Lifshitz black hole with two horizons (right panel).} \label{fig:WDWpatch}
\end{figure}

Following \cite{Lehner:2016vdi}, we introduce the null coordinates $u$ and $v$ by
\begin{equation}
u:= t+r^{\ast},\quad v:= t-r^{\ast},
\end{equation}
where $r^{\ast}=\int(\frac{l}{r})^{z+1}\frac{1}{f}dr$. Under the null coordinates the metric can be written as
\begin{equation}
ds^2 =-\left(\frac{r}{l}\right)^{2z}fdu^2 + 2\left(\frac{r}{l}\right)^{z-1}dudr + r^2 \sum\limits^{d}_{i=1}dx^{2}_{i},
\end{equation}
or
\begin{equation}
ds^2 =-\left(\frac{r}{l}\right)^{2z}fdv^2 + 2\left(\frac{r}{l}\right)^{z-1}dvdr + r^2 \sum\limits^{d}_{i=1}dx^{2}_{i}.
\end{equation}
For the choices of ($t,r$),($u,r$) and ($v,r$), we have
\begin{equation}
\int\sqrt{-g}d^{D}x=\Omega_{D-2}\int\frac{r^{D+z-3}}{l^{z-1}}drdw.
\end{equation}
where $w=\{t,\mu,\upsilon\}$. In this case, the WDW patch is shown in the left panel of Fig.~\ref{fig:WDWpatch}. As analyzed in the \cite{Alvarez:2014pra}, when $\beta_{1}\geq0,\beta_{2}\leq0$;
$\beta_{1}>0,\beta_{2}>0,1<z<D-2$ or $\beta_{1}<0,\beta_{2}<0,z>D-2$ , the nonlinearly charged Lifshitz black hole with exponent
 $z>1$ and $z\neq D-2$ has only one horizon.  To calculate the action growth, we consider the change of the on-shell action when the time has a shift from
$t_{0}$ to $t_{0}+\delta t$ on the left boundary, and as shown in Ref \cite{Chapman:2016hwi} that the joints at the UV cutoff surface and future singularity are unchanged
under the time translation. So we get
\begin{equation}
\delta I=I_{\mathcal{V}_1}-I_{\mathcal{V}_2}-\frac{1}{\kappa}\int_{S}Kd\Sigma+\frac{1}{\kappa}\int_{B'}adS-\frac{1}{\kappa}\int_{B}adS.
\end{equation}
With the equations of motion (\ref{equations of motion(2)}) the Lagrangian in the bulk has the form
\begin{equation}\label{bulk}
\mathcal{L}_{bulk}=\frac{-(z+D-2)+zM_{1}\left(\frac{l}{r}\right)^{D-2}-(D-2)M_{2}\left(\frac{l}{r}\right)^{z}}{\kappa l^{2}}.
\end{equation}
Thus the action in region $\mathcal{V}_{1}$ is
\begin{equation}
\begin{aligned}\label{V_{1}}
I_{\mathcal{V}_1}&=\frac{1}{2\kappa}l^{D-2}\Omega_{D-2}\int^{u_0+ \delta t}_{u_0} du \int_{\epsilon}^{{\rho}(u)}\left(\frac{r}{l}\right)^{z+D-3}\frac{-2(z+D-2)
+2zM_{1}\left(\frac{l}{r}\right)^{D-2}-2(D-2)M_{2}\left(\frac{l}{r}\right)^{z}}{l^{2}}dr\\
&=\frac{1}{2\kappa}l^{D-3}\Omega_{D-2}\int^{u_0+ \delta t}_{u_0} du\biggl[-2\left(\frac{r}{l}\right)^{z+D-2}+2M_{1}\left(\frac{r}{l}\right)^{D-2}-2M_{2}\left(\frac{r}{l}\right)^{z}\biggr]\bigg|^{\rho(u)}_{\epsilon}.
\end{aligned}
\end{equation}
Similarly, the contribution of $\mathcal{V}_{2}$ to the action is
\begin{equation}
\begin{aligned}
I_{\mathcal{V}_2}&=\frac{1}{2\kappa}l^{D-2}\Omega_{D-2}\int^{v_0+ \delta t}_{v_0} dv \int_{\rho_{1}(v)}^{{\rho_0}(v)}\left(\frac{r}{l}\right)^{z+D-3}\frac{-2(z+D-2)
+2zM_{1}\left(\frac{l}{r}\right)^{D-2}-2(D-2)M_{2}\left(\frac{l}{r}\right)^{z}}{l^{2}}dr\\
&=\frac{1}{2\kappa}l^{D-3}\Omega_{D-2}\int^{v_0+ \delta t}_{v_0} dv\biggl[-2\left(\frac{r}{l}\right)^{z+D-2}+2M_{1}\left(\frac{r}{l}\right)^{D-2}-2M_{2}\left(\frac{r}{l}\right)^{z}\biggr]\bigg|^{\rho_{0}(v)}_{\rho_{1}(v)}.
\end{aligned}
\end{equation}
Now we take the limit $\epsilon\rightarrow 0$, using a variables change: $u=u_{0}+v_{0}+\delta t-v$, then terms involving $\rho_{0}(u)$ and $\rho_{1}(v)$ cancel
out. So we get
\begin{equation}\label{V1-V2}
I_{\mathcal{V}_1}- I_{\mathcal{V}_2}=\frac{1}{2\kappa}l^{D-3}\Omega_{D-2}\int^{v_0+ \delta t}_{v_0} dv\biggl[-2\left(\frac{\rho_{1}}{l}\right)^{z+D-2}
+2M_{1}\left(\frac{\rho_{1}}{l}\right)^{D-2}-2M_{2}\left(\frac{\rho_{1}}{l}\right)^{z}\biggr].
\end{equation}
The variable $\rho_{1}(v)$ varies from $r_{B}$ to $r_{B'}$ as $v$ increases from $v_{0}$ to $v_{0}+\delta t$. But the variation is small, and one has
$r_{B'}=r_{B}+\mathcal{O}(\delta t)$ so that Eq. (\ref{V1-V2}) reduces to
\begin{equation}\label{V1-V2(2)}
I_{\mathcal{V}_1}- I_{\mathcal{V}_2}=\frac{1}{2\kappa}l^{D-3}\Omega_{D-2}\biggl[-2\left(\frac{r_{B}}{l}\right)^{z+D-2}+2M_{1}\left(\frac{r_{B}}{l}\right)^{D-2}-2M_{2}\left(\frac{r_{B}}{l}\right)^{z}\biggr] \delta t.
\end{equation}

Next we calculate the contribution from the spacelike surface $\mathcal{S}$ at $r=\epsilon$. The future-directed unit normal is given by
$n_{\alpha}=\frac{l}{r}|f|^{-\frac{1}{2}}\partial_{\alpha}r$, then the extrinsic curvature is
\begin{equation}\label{curvature}
K=\nabla_{\alpha}n^{\alpha}=-\frac{l^{z-1}}{r^{z+D-3}}\frac{d}{dr}\left(\frac{r^{z+D-2}}{l^{z}}|f|^{\frac{1}{2}}\right).
\end{equation}
Meanwhile the volume element is given by
\begin{equation}\label{approximating K}
d\Sigma=\Omega_{D-2}\frac{r^{z+D-2}}{l^{z}}|f|^{\frac{1}{2}}dt
\end{equation}
because $r=\epsilon\ll r_{+}$, approximatively $f\simeq-M_{1}\left(\frac{l}{r}\right)^{D-2}+M_{2}\left(\frac{l}{r}\right)^{z}$, one obtains
\begin{equation}\label{my case for boundary}
\begin{aligned}
I_{\mathcal{S}}&=-\frac{1}{\kappa}\int_{S} Kd\Sigma\\
&=\frac{1}{2\kappa }l^{D-3}\Omega_{D-2}\delta t\biggl[(-2z-D+2)M_{1}\left(\frac{r}{l}\right)^{z}+(z+2D-4)M_{2}\left(\frac{r}{l}\right)^{d-2}\biggr]\bigg|_{r=\epsilon}\\
&=0.
\end{aligned}
\end{equation}

Finally we calculate the contribution of the joint terms at $B,B^{'}$. Following \cite{Lehner:2016vdi}, the null joint rule states that
\begin{equation}\label{null joint}
a=\ln\left(-\frac{1}{2}k\cdot\overline{k}\right).
\end{equation}
Under affine parametrization the vectors $k^{\alpha}$ and $\overline{k}^{\alpha}$ can be expressed as
\begin{equation}\label{null joint}
k_{\alpha}=-c\partial_{\alpha}v=-c\partial_{\alpha}(t-r^{\ast}), \overline{k}_{\alpha}=\overline{c}\partial_{\alpha}u=\overline{c}\partial_{\alpha}(t+r^{\ast}),
\end{equation}
where $c$ and $\overline{c}$ are arbitrary positive constants. With these choices, we have $k\cdot\overline{k}=2c\overline{c}\frac{l^{2z}}{r^{2z}f}$, then we get
\begin{equation}
a=-\ln\left(-\frac{r^{2z}f}{l^{2z}c\overline{c}}\right).
\end{equation}
In the end we have
\begin{equation}
I_{B'B}=\frac{1}{2\kappa}\left(2\oint_{B^{'}}adS-2\oint_{B}adS\right)=\frac{1}{\kappa}\Omega_{D-2}[h(r_{B^{'}})-h(r_{B})],
\end{equation}
where $h(r):=-r^{D-2}\ln\left(-\frac{r^{2z}f}{l^{2z}c\overline{c}}\right)$. By making a Taylor expansion of $h(r)$ around $r=r_{B}$ and using
$dr=-\frac{1}{2}f\left(\frac{r}{l}\right)^{z+1}\delta t$, one obtains
\begin{equation}\label{my case for joint}
\begin{aligned}
I_{B'B}&=-\frac{1}{2\kappa}\Omega_{D-2}f\left(\frac{r}{l}\right)^{z+1}\frac{dh}{dr}\bigg|_{r=r_{B}}\delta t\\
&=\frac{\Omega_{D-2}}{2\kappa}\frac{r^{z+D-2}}{l^{z+1}}\biggl[2zf+r\frac{df}{dr}+(D-2)f\ln\left(-\frac{r^{2z}}{l^{2z}}\frac{f}{c\overline{c}}\right)\biggr]\bigg|_{r=r_{B}}\delta t.
\end{aligned}
\end{equation}
Collecting the formulae (\ref{V1-V2(2)}),(\ref{my case for boundary}) and (\ref{my case for joint}), and taking the late time limit $r_{B}\rightarrow r_{+}$ we have
\begin{eqnarray}
\label{late time complexity}
\frac{dI_{on-shell}}{dt}&=&\frac{1}{2\kappa}l^{D-3}\Omega_{D-2}\biggl\{\biggl[-2\left(\frac{r}{l}\right)^{z+D-2}+2M_{1}\left(\frac{r}{l}\right)^{D-2}-2M_{2}\left(\frac{r}{l}\right)^{z}\biggr]\nonumber\\
&&+\frac{r^{z+D-2}}{l^{z+D-2}}\biggl[2zf+r\frac{df}{dr}+(D-2)f\ln\left(-\frac{r^{2z}}{l^{2z}}\frac{f}{c\overline{c}}\right)\biggr]\bigg|_{r=r_{+}}\biggr\}\nonumber\\
&=&\frac{1}{2\kappa}l^{D-3}\Omega_{D-2}\biggl[(D-2)\left(\frac{r_{+}}{l}\right)^{z+D-2}+(D-2-z)M_{2}\left(\frac{r_{+}}{l}\right)^{D-2}\biggr].
\end{eqnarray}
Combining \eqref{charge potential} and \eqref{mass for my case}  we finally get
\begin{equation}\label{action growth}
\begin{aligned}
\frac{dI_{on-shell}}{dt}=\frac{z+D-2}{D-2}M-Q_{e}\Phi_{e}.
\end{aligned}
\end{equation}

Note that the WDW patch of charged Lifshitz black hole with a single horizon is similar to the one of neutral black hole as shown in Fig.~\ref{fig:WDWpatch}. Therefore the action growth rate of charged Lifshitz black hole with a single horizon \eqref{action growth} is also similar to the one of neutral black hole.  From \eqref{action growth} we see that, under the CA proposal in Lifshitz system, the late time action growth rate relates to the dynamics critical exponent $z$ and spacetime dimension $D$.
In \cite{An:2018xhv} the action growth rate also has an explicit dependence on $z$ and $d$, that is $dI_{on-shell}/dt=2(z+d-1)M/d$. In their case, there is a continuous limit as $z\rightarrow 1$ and  it reduces to the Ads black hole and we recovers the Lloyd bound $2M$ as $z=1$. In our case, however,  we don't have a continuous limit in the sense that we cannot choose the dynamics exponent $z=1$ because Proca field disappears when $z=1$. As a consequence in the limit $z\rightarrow 1$, the usual Lloyd bound $2M$ cannot be recovered from \eqref{action growth}. This implies more nontrivial features of our model and may provide more nontrivial informations about the validity of the CA duality as one applies it to test the Lloyd bound. In what follows we turn to this issue by examining whether the Lloyd bound is violated or not in our case.

\begin{figure}[t]
\begin{center}
\includegraphics[width=.90\textwidth]{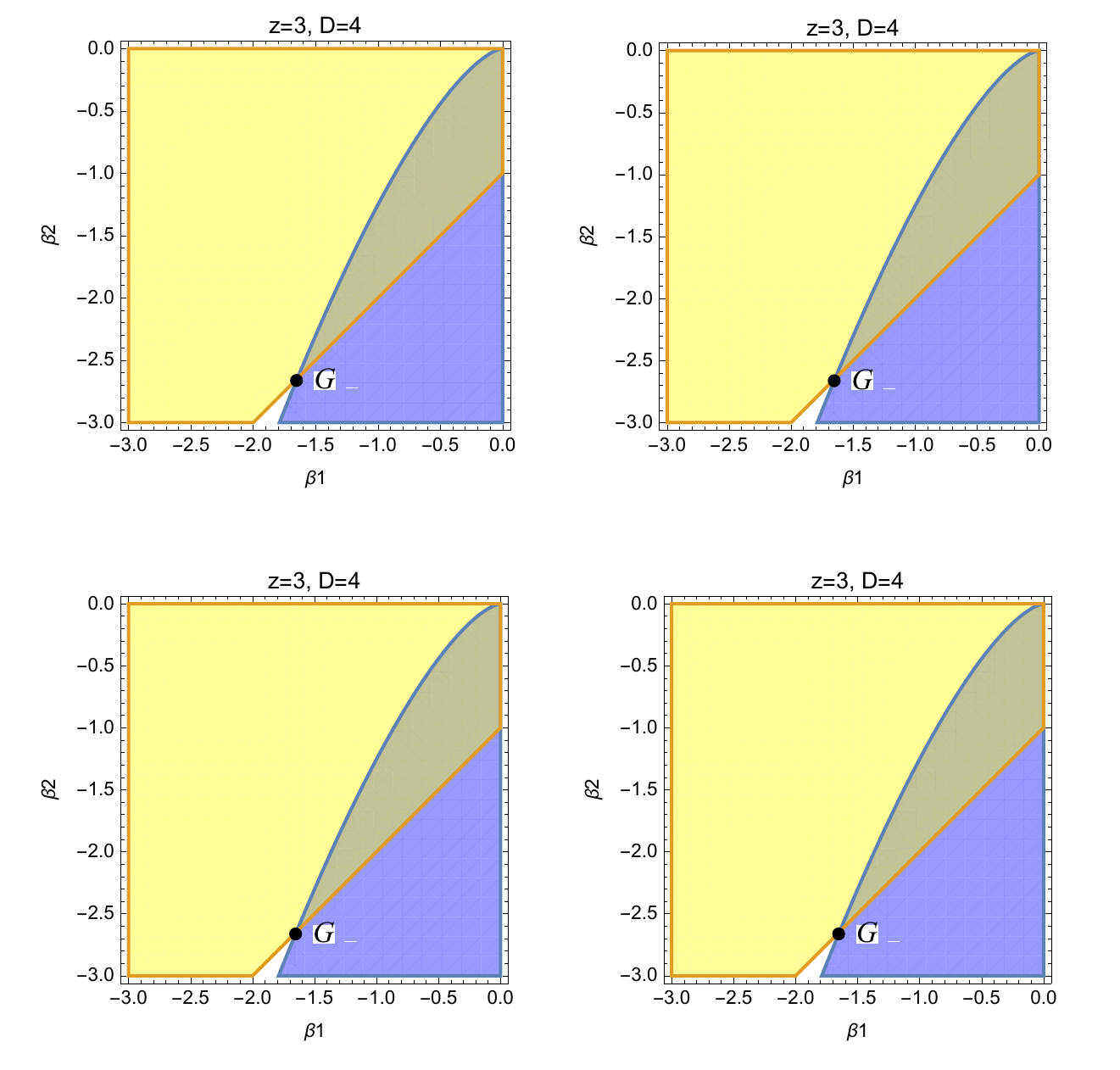}
\end{center}
  \caption{The case of
   $\beta_{1}<0,\beta_{2}<0,z>D-2$.  The blue part represent the regions where the the Lloyd bound is violated for any value of $\beta_{1},\beta_{2}$. The yellow part represent the limitations of $1>r_{+}/l>0$, imposing constraints  on the allowed values of $\beta_{1},\beta_{2}$  through the formula (\ref{gravitational potential}) for different $z$ and $D$. So, the overlapping part(grey part) in the parameter space is the regions where the Lloyd bound is violated. The point G is a critical point. It denotes the maximum that   $|\beta_{1}\beta_{2}|$ can take within the region where the Lloyd bound is violated.   }
  \label{fig:singlez=3}
\end{figure}

\begin{figure}[h]
\begin{center}
\includegraphics[width=.90\textwidth]{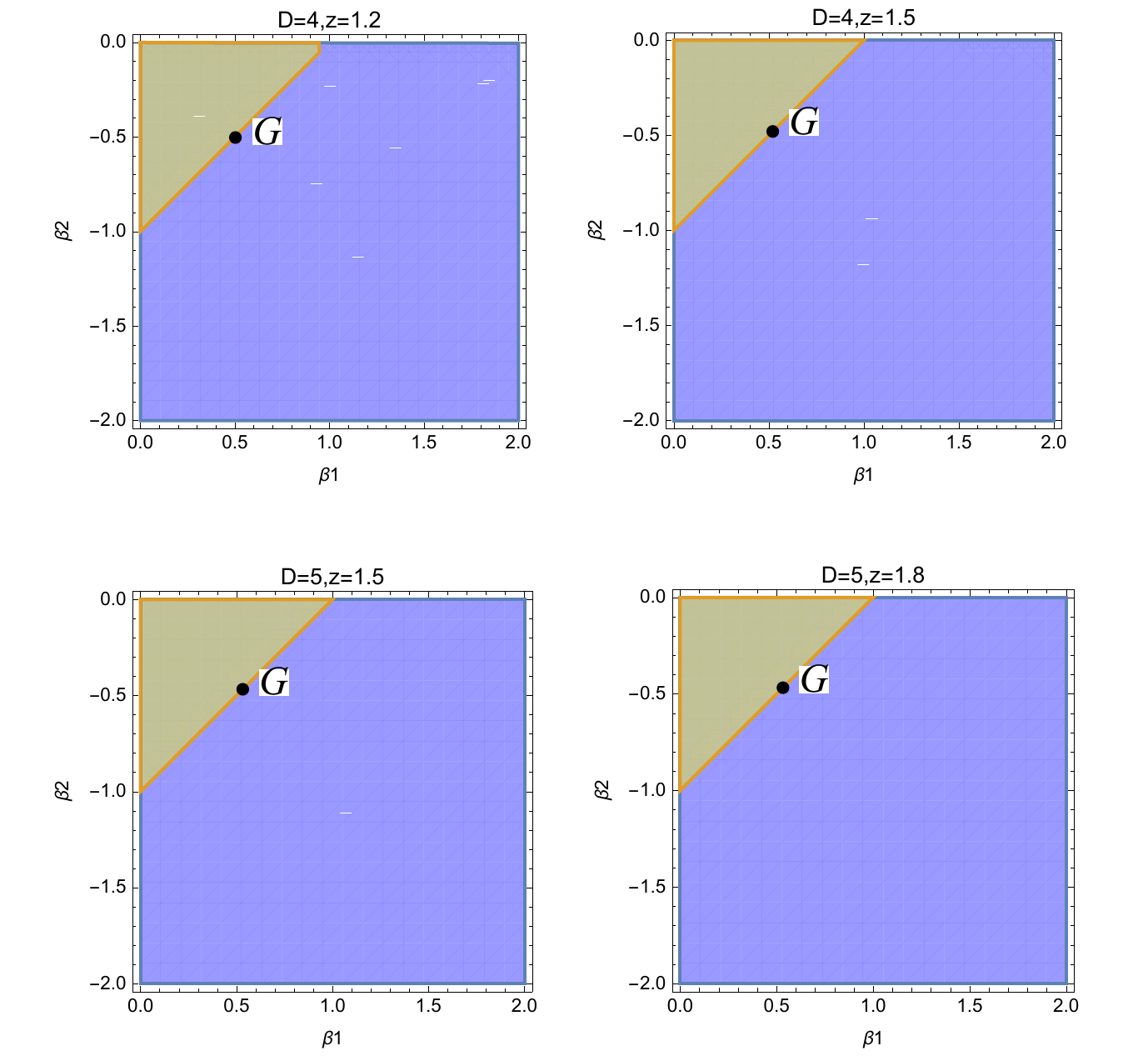}
\end{center}
  \caption{The case of
   $\beta_{1}>0,\beta_{2}<0,1<z<D-2$. The blue part represent the regions where the the Lloyd bound is violated for any value of $\beta_{1},\beta_{2}$. The yellow part represent the limitations of $1>r_{+}/l>0$, imposing constraints  on the allowed values of $\beta_{1},\beta_{2}$  through the formula (\ref{gravitational potential}) for different $z$ and $D$. So, the overlapping part(grey part) in the parameter space is the regions where the Lloyd bound is violated. The point G is a critical point. It denotes the maximum that  $|\beta_{1}\beta_{2}|$ can take within the region where the Lloyd bound is violated. }
  \label{fig:singlez=1p2}
\end{figure}

To proceed, let us first consider a special case where  $\beta_{1}=0$. In this case the action reduces to the power-law electrodynamic and the mass of black hole vanishes. Obviously, in order to have a well-defined horizon, $\beta_{2}$ must be less than zero and the action growth rate is $\frac{dI_{on-shell}}{dt}=-zM_{2}(\frac{r_{+}}{l})^{D-2}$, which indicates that the Lloyd bound is violated identically. In the same way, if we let $\beta_{2}=0$ so $\beta_{1}>0$, we find that the Lloyd bound is also violated.
Thermodynamics of these cases were studied in Refs \cite{Bravo-Gaete:2015xea,Dehghani:2013mba}. For the more general situation, when
$\beta_{1}>0,\beta_{2}<0,z>D-2$ and $\beta_{1}>0,\beta_{2}>0,1<z<D-2$ the Lloyd bound is still violated because the mass of the black hole is negative, but the action growth rate must be positive. It's worth to note that in both cases the black hole mass is negative.
 While for the case of $\beta_{1}<0,\beta_{2}<0,z>D-2$, the Lloyd bound is violated for part of the parameter space as shown in Fig.~\ref{fig:singlez=3}, and the range where the bound is violated increase as $z$ increases and decreases as $D$ increases. For the case of $\beta_{1}>0,\beta_{2}<0,1<z<D-2$ the bound is violated for all allowed value of $\beta_{1},\beta_{2}$ and it has no influence on the range where the bound is violated when we change the value of  $D$ and $z$ as shown in Fig.~\ref{fig:singlez=1p2} .


\subsection{ The case with two horizon}

As analyzed in the \cite{Alvarez:2014pra}, when $\beta_{1}>0,\beta_{2}>0,z>D-2$ or $\beta_{1}<0,\beta_{2}<0,1<z<D-2$ , the nonlinearly charged Lifshitz black hole
 with exponent $z>1$ and $z\neq D-2$ has two horizon, and the WDW patch is shown in the right panel of Fig.~\ref{fig:WDWpatch}. With the equation of motion
 (\ref{equations of motion(2)}), after taking the late-time limit, $I_{\mathcal{V}_{1}}-I_{\mathcal{V}_{2}}$ becomes
\begin{equation}\label{V1-V2(2)1}
I_{\mathcal{V}_{1}}-I_{\mathcal{V}_{2}}=\frac{1}{2\kappa}l^{D-3}\Omega_{D-2}\biggl[-2\left(\frac{r}{l}\right)^{z+D-2}+2M_{1}\left(\frac{r}{l}\right)^{D-2}-2M_{2}\left(\frac{r}{l}\right)^{z}\biggr] \bigg|^{r_{+}}_{r_{-}}\delta t.
\end{equation}
There are four joints contributions with joints $B^{'},B$ inside the past horizon and joints $C,C^{'}$ inside the future horizon. Then we have
\begin{equation}
\begin{aligned}\label{joint(1)}
I_{B'B}+I_{C'C}&=-\frac{1}{2\kappa}\Omega_{D-2}f\left(\frac{r}{l}\right)^{z+1}\frac{dh}{dr}\bigg|^{r_{+}}_{r_{-}}\delta t\\
&=\Omega_{D-2}\frac{1}{2\kappa}\frac{r^{z+D-2}}{l^{z+1}}\biggl[2zf+r\frac{df}{dr}+(D-2)f\ln\left(-\frac{r^{2z}}{l^{2z}}\frac{f}{c\overline{c}}\right)\biggr]\bigg|^{r_{+}}_{r_{-}}\delta t.
\end{aligned}
\end{equation}
Combining (\ref{V1-V2(2)1}) and (\ref{joint(1)}), we find
\begin{equation}
\begin{aligned}\label{late time complexity1}
\frac{dI_{on-shell}}{dt}&=\frac{1}{2\kappa}l^{D-3}\Omega_{D-2}\biggl\{\biggl[-2\left(\frac{r}{l}\right)^{z+D-2}+2M_{1}\left(\frac{r}{l}\right)^{D-2}-2M_{2}\left(\frac{r}{l}\right)^{z}\biggr]\\
&\ \ \ +\frac{r^{z+D-2}}{l^{z+D-2}}\biggl[2zf+r\frac{df}{dr}+(D-2)f\ln\left(-\frac{r^{2z}}{l^{2z}}\frac{f}{c\overline{c}}\right)\biggr]\biggr\}\bigg|^{r_{+}}_{r_{-}}\\
&=\frac{1}{2\kappa}l^{D-3}\Omega_{D-2}\biggl[(D-2)M_{1}\left(\frac{r}{l}\right)^{z}-zM_{2}\left(\frac{r}{l}\right)^{D-2}\biggr]\bigg|^{r_{+}}_{r_{-}}\\
&=T_{+}S_{+}-T_{-}S_{-}.
\end{aligned}
\end{equation}
Where $T_{-}(T_{+})$ and $S_{-}(S_{+})$ are the temperature and entropy of the inner(outer) horizon.  We find the late time action growth rate for the charged Lifshitz black hole with exponent $z>1$ and $z\neq D-2$ satisfies the universal expression in Refs \cite{Cai:2016xho}. Specifically, using  the Smarr formula (\ref{Smarr formula}) we can rewrite the growth rate of action as
\begin{equation}\label{action growth 22}
\frac{dI_{on-shell}}{dt}=\left(M-Q_{e}\Phi_{e}\right)_{+}-\left(M-Q_{e}\Phi_{e}\right)_{-}.
\end{equation}
This shows that the growth rate of the action for nonlinear charged Lifshitz black hole satisfies the universal expression in \cite{Cai:2016xho}.
In other words, the statement that the action growth bound is independent of the charged black hole size\cite{Cai:2016xho} is also valid for some cases of the Lifshitz system.

\section{ Action growth in nonlinearly charged Lifshitz black hole with exponent $z=D-2$}

In this section we would like to investigate the action growth rate of nonlinearly charged logarithmic Lifshitz black hole with exponent $z=D-2$ so as to see if there is violation of the action growth bound.

\subsection{ The case with a single horizon}

The logarithmic Lifshitz black hole has a single horizon when $\beta_{1}<0$. The Lagrangian now becomes
\begin{equation}\label{log bulk}
\mathcal{L}_{bulk}=\frac{1}{\kappa l^{2}}\biggl[-2(D-2)+M_{1}\left(\frac{l}{r}\right)^{D-2}-(D-2)M_{2}\left(\frac{l}{r}\right)^{D-2}+(D-2)\ln\left(\frac{r}{l}\right)M_{1}\left(\frac{l}{r}\right)^{D-2}\biggr].
\end{equation}
The on-shell action in region $\mathcal{V}_{1}$ is
\begin{equation}
\begin{aligned}\label{logV_{1}}
I_{\mathcal{V}_1}&=\frac{1}{2\kappa}l^{D-2}\Omega_{D-2}\int^{u_0+ \delta t}_{u_0} du \int_{\epsilon}^{{\rho}(u)}\left(\frac{r}{l}\right)^{2D-5}\biggl[-4(D-2)+2M_{1}\left(\frac{l}{r}\right)^{D-2}\\
&\ \ \ -2(D-2)M_{2}\left(\frac{l}{r}\right)^{D-2}+2(D-2)\ln\left(\frac{r}{l}\right)M_{1}\left(\frac{l}{r}\right)^{D-2}\biggr]dr\\
&=\frac{1}{2\kappa}l^{D-3}\Omega_{D-2}\int^{u_0+ \delta t}_{u_0} du\biggl[-2\left(\frac{r}{l}\right)^{2D-4}-2M_{2}\left(\frac{r}{l}\right)^{D-2}+2\ln\left(\frac{r}{l}\right)M_{1}\left(\frac{r}{l}\right)^{D-2}\biggr]\bigg|^{\rho(u)}_{\epsilon}.
\end{aligned}
\end{equation}

\begin{figure}[t]
\begin{center}
\includegraphics[width=.90\textwidth]{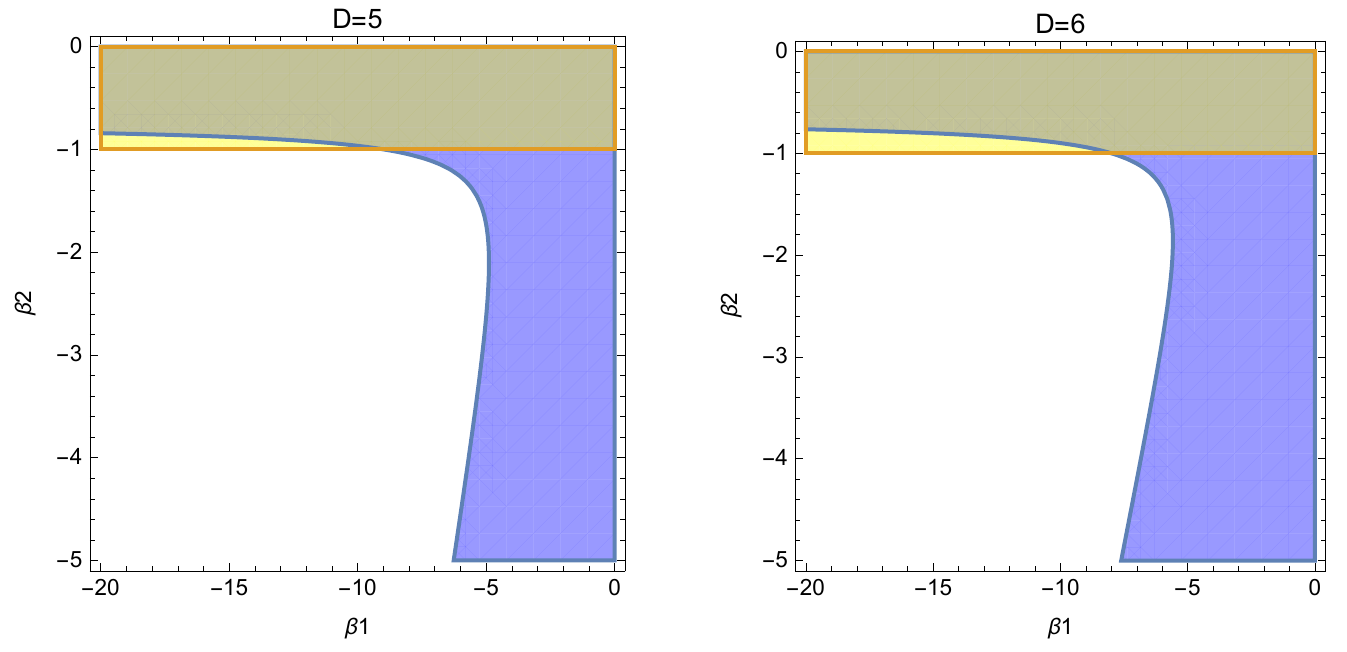}
\end{center}
  \caption{The case of
   $\beta_{1}<0,\beta_{2}<0$. The blue part represent the regions where the the Lloyd bound is violated for any value of $\beta_{1},\beta_{2}$. The yellow part represent the limitations of $1>r_{+}/l>0$, imposing constraints  on the allowed values of $\beta_{1},\beta_{2}$  through the formula (\ref{gravitational potential}) for different $z$ and $D$. So, the overlapping part(grey part) in the parameter space is the regions where the Lloyd bound is violated.}
  \label{fig:singlez=2}
\end{figure}

Similarly, the contribution of $\mathcal{V}_{2}$ to the action is
\begin{align}\label{logV_{2}}
I_{\mathcal{V}_2}=\frac{1}{2\kappa}l^{D-3}\Omega_{D-2}\int^{u_0+ \delta t}_{u_0} du\biggl[-2\left(\frac{r}{l}\right)^{2D-4}-2M_{2}\left(\frac{r}{l}\right)^{D-2}+2\ln\left(\frac{r}{l}\right)M_{1}\left(\frac{r}{l}\right)^{D-2}\biggr]\bigg|^{\rho_{0}(v)}_{\rho_{1}(v)}.
\end{align}
In the late time, (\ref{logV_{1}}) together with (\ref{logV_{2}}) lead to
\begin{equation}
\begin{aligned}\label{logV{1}-logV{2}}
I_{\mathcal{V}_1}-I_{\mathcal{V}_2}&=\frac{1}{2\kappa}l^{D-3}\Omega_{D-2}\delta t
\biggl[-2\left(\frac{r}{l}\right)^{2D-4}-2M_{2}\left(\frac{r}{l}\right)^{D-2}+2\ln\left(\frac{r}{l}\right)M_{1}\left(\frac{r}{l}\right)^{D-2}\biggr]\bigg|^{r_{+}}_{\epsilon}\\
&=0.
\end{aligned}
\end{equation}
The contribution of the spacelike surface $\mathcal{S}$ is
\begin{equation}\label{my case for log boundary}
\begin{aligned}
I_{\mathcal{S}}=\frac{l^{D-3}}{\kappa}\Omega_{D-2}\delta t\biggl[(2D-4)\left(\frac{r}{l}\right)^{2D-4}f+\frac{1}{2}\frac{r^{2D-3}}{l^{2D-4}}f'\biggr]\bigg|_{r=\epsilon}=0.
\end{aligned}
\end{equation}
The joint term is
\begin{equation}\label{my case for joint 1}
\begin{aligned}
I_{B'B}&=\frac{1}{2\kappa}\Omega_{D-2}\frac{r^{2D-4}}{l^{D-1}}rf'\bigg|_{r=r_{+}}\\
&=\frac{l^{D-3}}{2\kappa}\Omega_{D-2}\biggl\{(D-2)\left(\frac{r_{+}}{l}\right)^{D-2}\biggl[M_{1}\ln\left(\frac{r_{+}}{l}\right)-M_{2}\biggr]-\left(\frac{r_{+}}{l}\right)^{D-2}M_{1}\biggr\}
\end{aligned}
\end{equation}
Putting the term (\ref{logV{1}-logV{2}}),(\ref{my case for log boundary}) and (\ref{my case for joint 1}) together, we have
\begin{equation}
\begin{aligned}\label{dc/dt log}
\frac{dI_{on-shell}}{dt}&=\frac{l^{D-3}}{2\kappa}\Omega_{D-2}\biggl\{(D-2)\left(\frac{r_{+}}{l}\right)^{2(D-2)}-\left(\frac{r_{+}}{l}\right)^{D-2}M_{1}\biggr\}\\
&=2M-Q_{e}\Phi_{e}.
\end{aligned}
\end{equation}
It has the same form as the late time action growth rate (\ref{action growth}) if we let $z=D-2$.  This means that the logarithmic behavior doesn't influence the late time action growth rate for charged nonlinear Lifshitz black hole with a single horizon. On the other hand, although the requirement being a single horizon imposes a constraint on $\beta_1$ by $\beta_{1}<0$, there is no limitation on the value of $\beta_{2}$.
For the case of $\beta_{2}\geq0$, the Lloyd bound is violated and the black hole mass is less than or equal to zero.
For the case of $\beta_{2}<0$  the region where the Lloyd bound is violated  in the parameter space is shown in the Fig.~\ref{fig:singlez=2}. It shows that the range where the bound is violated decreases as $D$ increases.

\subsection{ The case with two horizon}

As $\beta_{1}>0$ the corresponding logarithmic charged Lifshitz black hole has two horizons.  It is not difficult to show that the bulk contribution is given by
\begin{equation}\label{logV{1}-logV{2}2}
I_{\mathcal{V}_1}-I_{\mathcal{V}_2}=\frac{1}{2\kappa}l^{D-3}\Omega_{D-2}\delta t
\biggl[-2\left(\frac{r}{l}\right)^{2D-4}-2M_{2}\left(\frac{r}{l}\right)^{D-2}+2\ln\left(\frac{r}{l}\right)M_{1}\left(\frac{r}{l}\right)^{D-2}\biggr]\bigg|^{r_{+}}_{r_{-}}.
\end{equation}
As shown in Fig.~\ref{fig:WDWpatch}, four joints $B^{'},B$ and $C,C^{'}$ also have contributions to the action, which totally is given by
\begin{equation}\label{my case for joint 2}
\begin{aligned}
I_{B'B}+I_{C'C}&=\frac{1}{2\kappa}\Omega_{D-2}\frac{r^{2D-4}}{l^{D-1}}rf'\bigg|_{r_{-}}^{r_{+}}\\
&=\frac{l^{D-3}}{2\kappa}\Omega_{D-2}\biggl\{(D-2)\left(\frac{r}{l}\right)^{D-2}\biggl[M_{1}\ln\left(\frac{r}{l}\right)-M_{2}\biggr]-\left(\frac{r}{l}\right)^{D-2}M_{1}\biggr\}\bigg|_{r_{-}}^{r_{+}}.
\end{aligned}
\end{equation}
Combining (\ref{logV{1}-logV{2}2}) and (\ref{my case for joint 2}), we have
\begin{equation}
\begin{aligned}
\frac{dI_{on-shell}}{dt}&=\frac{l^{D-3}}{2\kappa}\Omega_{D-2}\biggl\{(D-2)\left(\frac{r_{+}}{l}\right)^{D-2}\biggl[M_{1}\ln\left(\frac{r_{+}}{l}\right)-M_{2}\biggr]-\left(\frac{r_{+}}{l}\right)^{D-2}M_{1}\biggr\}\bigg|_{r_{-}}^{r_{+}}\\
&=T_{+}S_{+}-T_{-}S_{-}.
\end{aligned}
\end{equation}
Using  the Smarr formula (\ref{Smarr formula}), we can obtain the growth rate of action similar as the one for the AdS-RN black hole \cite{Brown:2015lvg,Cai:2016xho}
\begin{equation}\label{action growth 2}
\frac{dI_{on-shell}}{dt}=(M-Q_{e}\Phi_{e})_{+}-(M-Q_{e}\Phi_{e})_{-}.
\end{equation}
This means that the logarithmic behavior has no effect on the late time action growth rate of the nonlinear charged Lifshitz black holes with two horizons.

\section{Conclusions and discussions}

In this paper, we use the CA conjecture to investigate the late time complexity growth rate of nonlinearly charged Lifshitz black holes and nonlinearly charged logarithmic Lifshitz black holes with a single horizon and two horizons. In these cases, Proca field is the essential structure of the anisotropic Lifshitz vacuum. Since the mass of the Proca field is determined  by the dynamical exponent $z$ and spacetime dimension $D$ through (\ref{proca field}), $z$ and $D$ could precisely reflect the effect of the Proca field to the holographic complexity. Our results show that the value of dynamics exponent $z$, spacetime dimension $D$ and dimensionless coupling constants $\beta_{1},\beta_{2}$ will influence the late time action growth rate (\ref{late time complexity}). Comparing the nonlinearly charged Lifshitz black holes and those logarithmic ones, we find although on the side of gravity when $z=D-2$ the Lifshitz black hole will yields logarithmic decay behavior, but on the side of holographic complexity it cannot yield any specific influence.

\begin{table}
\caption{The effect of different parameter choices on the Lloyd bound}
\label{tab:all cases}
\begin{center}
\begin{tabular}{|l|l|l|l|}
\hline
Dynamic exponent & Dimensionless constants  & Black hole mass & Lloyd bound \\
\hline
$z>1$ and $z\neq D-2$  & $\beta_{1}=0,\beta_{2}<0$ & zero & violated \\
$z>1$ and $z\neq D-2$  & $\beta_{1}>0,\beta_{2}=0$ & zero & violated \\
$z>D-2$ & $\beta_{1}>0,\beta_{2}<0$ & negative & violated \\
$1<z<D-2$ & $\beta_{1}>0,\beta_{2}>0$ & negative & violated \\
$z>D-2$ & $\beta_{1}<0,\beta_{2}<0$ & positive & partly violated as shown in the Fig.~\ref{fig:singlez=3} \\
$1<z<D-2$ & $\beta_{1}>0,\beta_{2}<0$ & positive & violated \\
$z=D-2$ & $\beta_{1}<0,\beta_{2}=0$ & zero & violated \\
$z=D-2$ & $\beta_{1}<0,\beta_{2}>0$ & negative & violated \\
$z=D-2$ & $\beta_{1}<0,\beta_{2}<0$ & positive & partly violated as shown in the Fig.~\ref{fig:singlez=2} \\
\hline
\end{tabular}
\end{center}
\end{table}

\begin{figure}[h]
\begin{center}
\includegraphics[width=.90\textwidth]{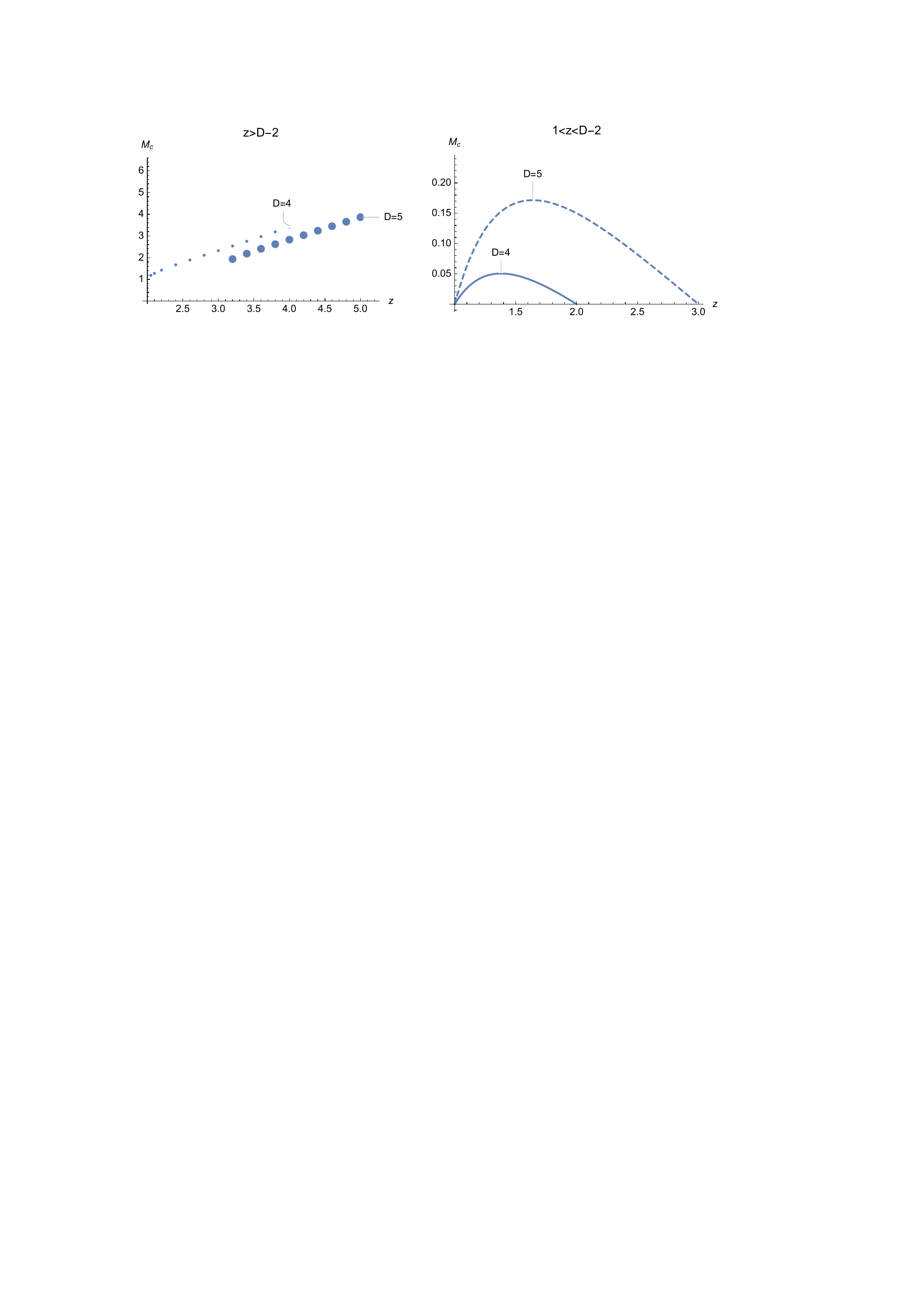}
\end{center}
  \caption{ The critical mass $M_c$ as a function of  $z$ for $D=4$ and $D=5$. Left panel is the case of $z>D-2$. Right panel is the case of $1<z<D-2$.   }
  \label{Mc}
\end{figure}

The other result of this paper is that the action growth bound does not always hold for the present model. More specifically,  for the black hole with two horizons, the action growth bound is always valid. While for the black hole with a single horizon, whether the bound hold depends on the different values of parameters. The details are listed in the Table.~\ref{tab:all cases}. It seems that black hole mass plays an important role in the action growth rate. More specifically, for black hole with negative mass, the Lloyd bound is identically violated. For positive-mass black hole with $z>1$ and $z\neq D-2$, however, we find that black holes with smaller mass are more likely to violate the Lloyd bound. Actually, from Figs.~\ref{fig:singlez=3} and \ref{fig:singlez=1p2} we see that there is a critical point (the point G). It denotes the maximal value that $|\beta_1\beta_2|$ can take to let the Lloyd bound be violated. It relates to the black hole mass through definition \eqref{mass for my case} and \eqref{integration constant}. In this way we can define a critical mass (denoted by $M_{c}$) of black hole, above which there is no violation of the Lloyd bound.
We discretely compute the critical mass $M_c$ for different $z$ when $z>D-2$. We plot them as a function of $z$ as shown in the left panel of Fig.~\ref{Mc}. For the case of $1<z<D-2$, the critical point G are same for different $z$ and $D$. The relationship between the critical mass $M_c$ and $z$ now can be ploted as a continuous function as shown in the right panel of Fig.~\ref{Mc}. For the positive mass black hole with $z=D-2$, however, there is no obvious relationship between the violation of the Lloyd bound and the black hole mass (\ref{mass for my case log}). Although $|\beta_{2}|$ can only choose small values ($|\beta_2|<1$), there is no limitation on the value of $|\beta_{1}|$. Therefore we could get black hole with arbitrary mass such that the Lloyd bound is still violated as shown in the Fig.~\ref{fig:singlez=2}. In other words, there is no critical mass for this case.

It would be very interesting to check whether the Lloyd bound is still violated for those values of parameters when we use the CV conjecture to our model. Especially,  there is an improved version of the CV conjecture---the``complexity=volume 2.0" (CV 2.0) --- which was proposed in \cite{Couch:2016exn}. It was found that the CV 2.0 would not violate the Lloyd bound for many cases where the Lloyd bound is violated in the framework of the CA conjecture. The same thing deserves to do for the so-called CA2.0, which is a modified version of CA conjecture proposed in\cite{Fan:2018wnv}. Very more recently, there are two more versions of CV conjecture  \cite{Liu:2019mxz,Wang:2019zqw} which show that the Lloyd bound holds under their framework even for the cases where the bound is violated in the original CA or CV conjecture. We expect the system considered in the present paper, due to its non-relativistic  and nonlinear nature, provides a very good probe to test their validity or rationality of these improved conjectures. We leave these investigations to the future study.

\section*{\bf Acknowledgements}This work was supported in part by the National Natural Science Foundation of China under grant numbers 11975116, 11665016, 11565019 and 11563006, and Jiangxi Science Foundation for Distinguished Young Scientists under grant number 20192BCB23007.

\end{document}